\documentclass{ifacconf}

\newtheorem{assumption}{Assumption}
\newtheorem{remark}{Remark}
\newtheorem{lemma}{Lemma}
\newtheorem{definition}{Definition}
\newtheorem{theorem}{Theorem}[section]
\usepackage{graphicx}      
\usepackage{natbib}        
\usepackage{amsmath,amssymb,array,arydshln,color}
\usepackage{multirow}
\usepackage{diagbox}
\usepackage{subcaption}
\usepackage{caption}
\usepackage{algpseudocode}
\usepackage{algorithm}
\usepackage{tabularx}
\usepackage[colorinlistoftodos]{todonotes}
\makeatletter
\newcommand{\multiline}[1]{%
  \begin{tabularx}{\linewidth-\ALG@thistlm-0.4cm}[t]{@{}X@{}}
    #1
  \end{tabularx}
}
\makeatother

\def\Ac{{\mathcal A}}

\def\Cbb{{\mathbb C}}

\def\Dc{{\mathcal D}}

\def\Ec{{\mathcal E}}

\def\Gc{{\mathcal G}}

\def\Lc{{\mathcal L}}

\def\Nc{{\mathcal N}}

\def\pfr{{\mathfrak p}}

\def\Qc{{\mathcal Q}}

\def\qfr{{\mathfrak q}}

\def\Rbb{{\mathbb R}}

\def\Vc{{\mathcal V}}

\def\0{{\bf 0}}

\newcommand{\bitem}{\begin{itemize}}
\newcommand{\eitem}{\end{itemize}}
\newcommand{\btabular}{\begin{tabular}}
\newcommand{\etabular}{\end{tabular}}
\newcommand{\bcenter}{\begin{center}}
\newcommand{\ecenter}{\end{center}}
\newcommand{\bea}{\begin{eqnarray}}
\newcommand{\eea}{\end{eqnarray}}
\newcommand{\bean}{\begin{eqnarray*}}
\newcommand{\eean}{\end{eqnarray*}}
\newcommand{\ba}{\left[ \begin{array}}
\newcommand{\ea}{\\ \end{array} \right]}
\newcommand{\bear}{\begin{array}}
\newcommand{\eear}{\\ \end{array}}

\newcommand{\non}{\nonumber}

\newcommand*{\QEDB}{\hfill\ensuremath{\blacksquare}}%
\newcommand*{\QET}{\hfill\ensuremath{\triangleleft}}
\newcommand{\norm}[1]{\left\lVert#1\right\rVert}

\newcounter{subequation}
\def\beasub{\addtocounter{equation}{+1}
\setcounter{subequation}{\value{equation}}
\setcounter{equation}{0}
\renewcommand{\theequation}{\arabic{subequation}\alph{equation}}
\begin{eqnarray}}
\def\eeasub{\end{eqnarray}
\setcounter{equation}{\value{subequation}}
\renewcommand{\theequation}{\arabic{equation}}}

\def\inf{\operatornamewithlimits{inf\vphantom{p}}}
\newtheorem{problem}{Problem}

\begin{document}
\begin{frontmatter}

\title{Optimal Detector Placement in Networked Control Systems under Cyber-attacks with Applications to Power Networks\thanksref{footnoteinfo} 
} 
\thanks[footnoteinfo]{This work is supported by the Swedish Research Council under
the grants 2018-04396 and 2021-06316 and by the Swedish Foundation for Strategic Research.}

\author[IT]{Anh Tung Nguyen},
\author[EE]{Sribalaji C. Anand},
\author[IT]{Andr{\'e} M. H. Teixeira},
\author[IT]{Alexander Medvedev}

\address[IT]{Department of Information Technology, Uppsala University, PO Box 337, SE-75105, Uppsala, Sweden (e-mail: \{anh.tung.nguyen, andre.teixeira, alexander.medvedev\}@it.uu.se)}
\address[EE]{Department of Electrical Engineering, Uppsala University, PO Box 65, SE-75103, Uppsala, Sweden (e-mail: sribalaji.anand@angstrom.uu.se)}

\begin{abstract}   
This paper proposes a game-theoretic method to address the problem of optimal detector placement in a networked control system under cyber-attacks.
The networked control system is composed of interconnected agents where each agent is regulated by its local controller over unprotected communication, which leaves the system vulnerable to malicious cyber-attacks.
To guarantee a given local performance, the defender optimally selects a single agent on which to place a detector at its local controller with the purpose of detecting cyber-attacks.
On the other hand, an adversary optimally chooses a single agent on which to conduct a cyber-attack on its input with the aim of maximally worsening the local performance  while remaining stealthy to the defender. 
First, we present a necessary and sufficient condition to ensure that the maximal attack impact on the local performance is bounded, which restricts the possible actions of the defender to a subset of available agents.
Then, by considering the maximal attack impact on the local performance as a game payoff, we cast the problem of finding optimal actions of the defender and the adversary as a zero-sum game.
Finally, with the possible action sets of the defender and the adversary, an algorithm is devoted to determining the Nash equilibria of the zero-sum game that yield the optimal detector placement.
The proposed method is illustrated on an IEEE benchmark for power systems.
\end{abstract}


\end{frontmatter}

\section{Introduction}
Society's rising demands require the development of complex and networked systems such as power grids, transportation networks, and water distribution networks.
To enhance the performance and the efficiency of such systems, they might be divided into interconnected subsystems which are managed remotely through insecure  communication channels.
This insecure protocol possibly leaves the networked control systems vulnerable to cyber-attacks such as false data injection, covert, and replay attacks \citep{teixeira2015secure}, inflicting serious civil damages and financial loss. 
In the last decade, an Iranian industrial control system and a Ukrainian power grid have witnessed the catastrophic consequences of malware such as Stuxnet \citep{falliere2011w32} and Industroyer \citep{kshetri2017hacking}, respectively.
Motivated by the above observation, defense strategies are needed to deal with such cyber-attacks with the purpose of protecting the networked control systems.

In this paper, we deal with the problem of optimal detector placement against a cyber-adversary in a networked control system which is represented by interconnected linear second-order agents. 
Every agent is regulated by its local controller through unprotected communication, which leaves the system vulnerable to malicious cyber-attacks.
To guarantee a given local performance, the defender selects an agent on which to place a detector at its controller with the purpose of detecting malicious cyber-attacks.
Meanwhile, the malicious adversary chooses an agent on which to inject attack signals with the purpose of maximally worsening the local performance while remaining stealthy to the defender.
The contributions of this paper are the following: 
\begin{enumerate}
    \item The boundedness of the worst-case attack impact is guaranteed by a necessary and sufficient condition based on the suitable choices of control parameters and the system-theoretic property of the underlying dynamical system, namely relative degree.
    This condition restricts the possible choices of the defender to a subset of available agents. 
    \item The bounded worst-case attack impact is employed as a game payoff that enables us to translate the purposes of the defender and the adversary into a zero-sum game.
    \item Based on the notions of the Nash equilibria \citep{zhu2015game}, an algorithm is devoted to determining Nash equilibria of the zero-sum game that yield the best strategies of the defender and the adversary.
\end{enumerate}
To illustrate the obtained results, we apply our proposed method to the IEEE 14-bus system which represents a portion of the American Power Network. 
We conclude this section by providing the notation used in this paper.
\\
{\bf Notation:}
the sets of real positive (negative) numbers are denoted as $\Rbb_{+}~(\Rbb_-)$; $\Rbb^n~(\Cbb^n)$ 
stands for sets of real (complex) $n$-dimensional vectors; every vector $v$ and matrix $A$ can be denoted $v = [v_i]$ where $v_i$ is  $i$-th element and $A = [a_{ij}]$ where $a_{ij}$ is $(i,j)$ entry, respectively;
$I$ stands for an identity matrix with an appropriate dimension.
Let us define $e_i \in \Rbb^n$ with all zero elements except the $i$-th element that is set as $1$.
Consider the norm {$\norm{x}_{\Lc_2 [0,T]}^2 \triangleq \frac{1}{T}\int_{0}^{T} \norm{x(t)}_2^2~dt$, where we simplify the notation to $\norm{x}^2_{\Lc_2}$ if the time horizon $[0,T]$ is clear from the context}.
The space of square-integrable  functions is defined as $\Lc_{2} \triangleq \bigl\{f: \Rbb_{+} \rightarrow \Rbb ~|~ \norm{f}^2_{\Lc_2 [0,\infty]} < \infty \bigr\} $
and the extended space be defined as $\Lc_{2e} \triangleq \bigl\{ f: \Rbb_{+} \rightarrow \Rbb ~|~ \norm{f}^2_{\Lc_2 [0,T]} < \infty,~ \forall~ 0 < T < \infty \bigr\} $.
Let $\Gc \triangleq (\Vc, \Ec, \Ac)$ be a graph with the set of $N$ vertices $\Vc = \{1, 2,\ldots,N\}$,
the set of edges $\Ec \subseteq \Vc \times \Vc $, and the  adjacency matrix $\Ac = [a_{ij}]$.
For every $(i,j \neq i) \in \Ec$,  $a_{ij} > 0$  
and with $(i,j) \notin \Ec$ or $i = j$, $a_{ij} = 0$. 
The degree of vertex $i$ is denoted as 
$\Delta_i =  \sum_{j=1}^{n} a_{ij}$ and the degree matrix of graph $\Gc$ is defined as 
$\Delta = {\bf diag}\big([\Delta_i]\big)$, where ${\bf diag}$ stands for a diagonal matrix.
The Laplacian matrix is defined as $L = [\ell_{ij}] = \Delta - \Ac$.
Further, $\Gc$ is called an undirected connected graph if and only if matrix $\Ac$ is symmetric and the algebraic multiplicity of zero as an eigenvalue of $L$ is one.
The set of all neighbours of vertex $i$ is denoted as $\Nc_i = \{j \in \Vc | (i,j) \in \Ec \}$.
We denote a set $\Vc_{-i} \triangleq \Vc \setminus \{i\}$.
\section{Problem formulation} 
This section first describes a networked control system under cyber-attacks.
Then, we introduce the resources and the strategies of the defender and the adversary.
Finally, the worst-case attack impact on the local performance is analyzed. 
\subsection{Networked control system  under cyber-attacks}
Consider an undirected connected  graph $\Gc \triangleq (\Vc, \Ec, \Ac)$ consisting of 
$N$ agents where each agent $i$ has a second-order state-space model:
\begin{align}
    \dot p_i(t) &= v_i(t), \label{sys:pi} \\
    m_i \dot v_i(t) &= \displaystyle  \sum_{j \in \Nc_i} \ell_{ij} \Big( p_i(t) - p_j(t) \Big)  - h_i v_i(t)  + \tilde u_i(t), \label{sys:vi} \\
    y_i(t) &= p_i(t), \label{sys:yi}
\end{align}
where $p_i(t),~v_i(t) \in \Rbb$ are the states, $\tilde u_i(t) \in \Rbb$ is the healthy/attacked input, and $y_i(t) \in \Rbb$ is the output  of agent $i$.
The local performance of the entire network is evaluated via the output energy over a given, possibly infinite, time horizon of a given agent $\rho \in \Vc$ denoted as $\norm{y_\rho}^2_{\Lc_{2}}$.
Parameters $m_i, h_i \in \Rbb_+$ and $\forall (i,j) \in \Ec, ~\ell_{ij} \in \Rbb_-$ are given.
We {utilize} the following healthy local control law, {which is adapted from \citet[Ch. 4]{tegling2018fundamental},} for each agent $i \in \Vc$
\begin{align}
    u_i(t) &=  - \theta_i y_i(t) + \phi_i \xi_i(t), \label{sys:ui} \\
    \dot \xi_i(t) &= -\frac{1}{\tau} \xi_i(t) - \frac{\kappa_D}{\tau} \dot y_i(t),
    \non
\end{align}
where $\xi_i(t)$ is a virtual control input of agent $i$ and $\theta_i, \phi_i, \kappa_D, \text{and}~ \tau \in \Rbb_+$ are control parameters.
If the communication channel to agent $i$ from its local controller is attacked by an adversary, $\tilde u_i(t) \neq u_i(t)$ which will be described in the following subsection; otherwise $\tilde u_i(t) = u_i(t)$.
Let us employ the following assumption.
\begin{assumption}
    \label{assumption:per_protect}
    The communication between the controller and the system of the given performance agent $\rho \in \Vc$ is protected from any cyber-attacks. 
    Further, its controller is unavailable for the defender to place a detector.
    \QET
\end{assumption}
For convenience, let us use the following notation in the remainder of the paper:
$p(t) \triangleq [p_i(t)]$, $v(t) \triangleq [v_i(t)]$,
$\xi(t) \triangleq [\xi_i(t)]$, 
$x(t) \triangleq [x_1(t)^\top,x_2(t)^\top,\ldots,x_N(t)^\top]^\top$ where $x_i(t) \triangleq [p_i(t),v_i(t),\xi_i(t)]^\top$,
$M \triangleq \textbf{diag}\big( [m_i] \big)$, $H \triangleq \textbf{diag}\big( [h_i] \big)$, $\Theta \triangleq \textbf{diag}\big( [\theta_i] \big)$, and $\Phi \triangleq \textbf{diag}\big( [\phi_i] \big)$.
{   
\begin{remark}
     The control law \eqref{sys:ui} will drive the system dynamics \eqref{sys:pi}-\eqref{sys:vi} to a closed-loop system that is different from the one in \citet[Ch. 4]{tegling2018fundamental}, due to no interaction of states $v_i$ among agents. Thus, we will need to show how this control law stabilizes the system \eqref{sys:pi}-\eqref{sys:vi} in Section~\ref{subsec:strategy}. Further, this control law plays an important role in the strategy of the defender which will be introduced in Section~\ref{sec:opt_dectector}.
\end{remark}
}
{
\begin{remark}
    In this study, we determine the local performance of the entire network  through the energy of the output measurement  of the  agent $\rho$ over a  possibly infinite time horizon. On the other hand, other local performances can be utilized based on different applications. We leave the comparison among local performances for future work.  
\end{remark}
}
\label{subsec:resource}
\subsection{Resources of the adversary and the defender}
\subsubsection{\textbf{System knowledge:}}
The malicious adversary and the defender know the location of the given protected performance agent $\rho$, the appearance of their competitors, the agent set $\Vc$, and the edge set $\Ec$.
They also know all the system parameters $M$, $H$, $\Theta$, $\Phi$, $\kappa_D$, and $\tau$ as well as the detection mechanism which the defender will utilize.
\subsubsection{\textbf{Players' possible actions:}}
\begin{figure} [!t]
    \centering
    \includegraphics[width=\linewidth]{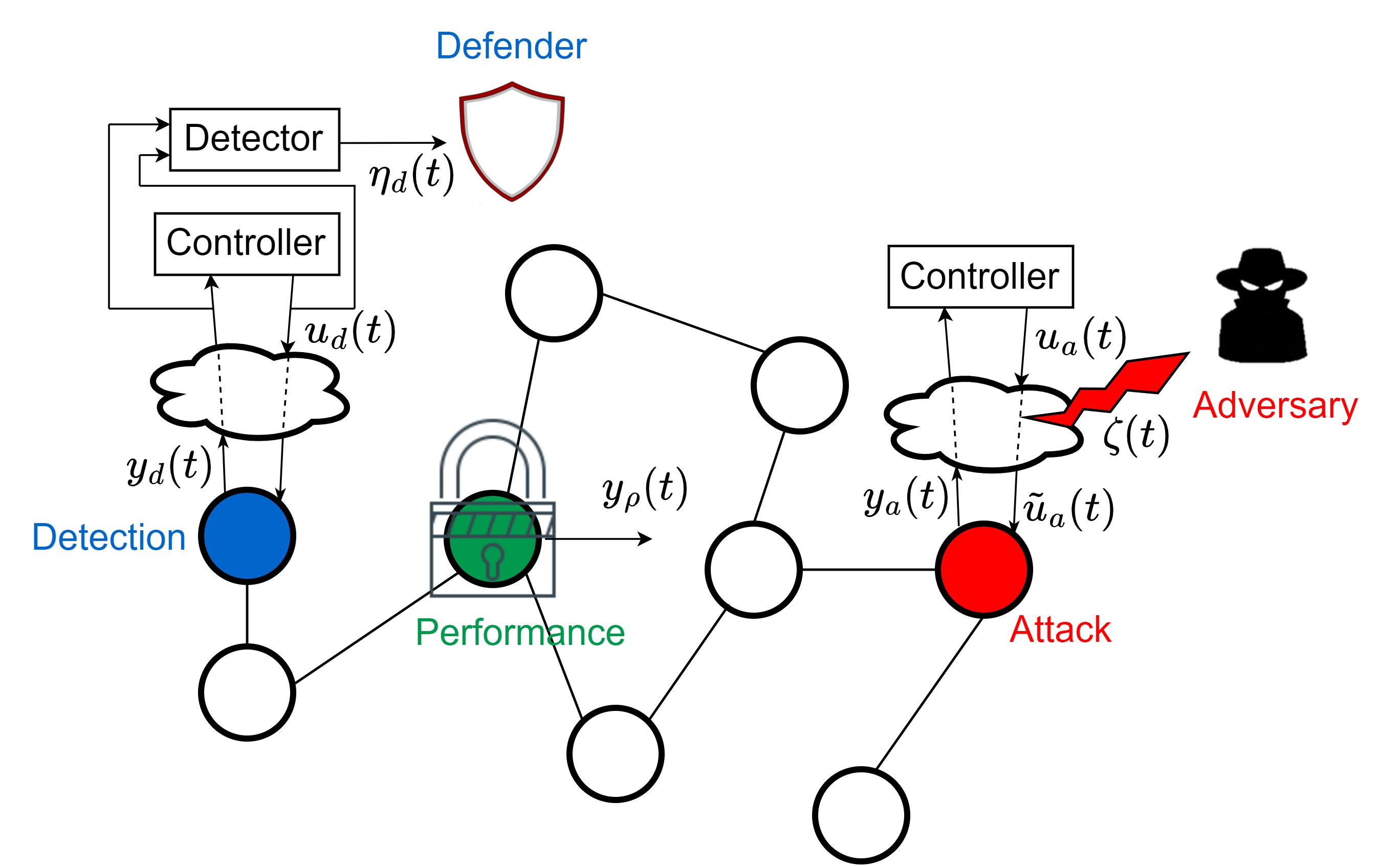}
    \caption{Illustration of a networked control system with the (green) protected performance agent under cyber-attack. While the defender selects the (blue) detection agent on which to place a detector, the adversary chooses the (red) attack agent on which to conduct a cyber-attack.}
    \label{fig:NCS_attack}
\end{figure}

According to \textit{Assumption~\ref{assumption:per_protect}},
each player is able to choose a single agent in $\Vc_{-\rho}$ to implement their strategies.
The adversary
selects the attack agent $a \in \Vc_{-\rho}$ on which to conduct a malicious attack signal $\zeta(t) \in \Rbb$ on its input with the aim of maximally disrupting the output of the  performance agent $\rho$ as follows:
\begin{align}
    \tilde u_i(t) = u_i(t) + \begin{cases}
        0,~~~ & i \in \Vc_{-a}, \\
        \zeta(t), & i \equiv a.
    \end{cases} \label{sys:uai}
\end{align}
Meanwhile, the defender chooses the detection agent $d \in \Vc_{-\rho}$ on which to place a detector that generates a residual signal with the purpose of detecting the cyber-attack.
These strategies of the two players are illustrated in Fig.~\ref{fig:NCS_attack} and described in detail below.
{
\begin{remark}
    In the scope of this study, we assume that the location of the performance agent $\rho$ is revealed to both the defender and the malicious adversary to simplify the security problem. 
    The problem of an unknown performance agent is left for future work.
\end{remark}
}
\subsection{Strategies of the adversary and the defender}
\label{subsec:strategy}
Before going into those strategies, let us rewrite the
closed-loop networked control system with its dynamics \eqref{sys:pi}-\eqref{sys:yi} under the control law \eqref{sys:ui}-\eqref{sys:uai} as follows
\begin{align}
    \dot x(t) &= A x(t) + E_a \zeta(t),
    \label{sys_cl:x} \\
    y_{i}(t) &= C_i x(t), \quad \forall i\in \mathcal{V}, \label{sys_cl:yi}\\
   {y_\rho(t)} &=  C_\rho x(t), \label{sys_cl:yrho}
\end{align}
where
\begin{align*}
    A &= \ba{ccc}
    0 &~~ I & ~~ 0 \\ 
    -M^{-1}(L+\Theta) ~~&~~ - M^{-1} H ~~&~~ M^{-1} \Phi \\  
    0 &~~ \displaystyle -\frac{\kappa_D}{\tau} I & -\frac{1}{\tau} I
    \ea, \\
    E_a &= \left[
    0^\top,~e_a^\top,~0^\top \right]^\top ,~
    C_i = \left[ e_i^\top,~0^\top,~0^\top \right]. 
\end{align*}
\begin{lemma}
\label{lem:A_hurwitz}
     Consider the healthy system \eqref{sys_cl:x} where $\zeta(t) = 0$ and assume that $\Gc$ is an undirected connected graph.
     Then, the matrix $A$ in \eqref{sys_cl:x} is Hurwitz.
     \QET
\end{lemma}
\begin{pf}
     Consider the candidate Lyapunov  function
     \begin{align}
         V\big(x(t)\big) = x(t)^\top \bar{P} x(t),
         \label{pf:lyap}
     \end{align}
     where 
     \begin{align}
         \bar{P} &= \ba{ccc}
         M^{-1}\big( L + \Theta + \sigma H \big)  ~~&~~ \sigma I ~~& 0 \\
         \sigma I & I & 0 \\
         0 & 0 &~~ \kappa_D^{-1}\tau M^{-1} \Phi 
         \ea, \non \\ 
         0 &< \sigma <  \min \Big\{ {\min_{i \in \Vc}}~ \frac{h_i}{m_i} ,~ \frac{4~ {\min_{i \in \Vc}}~ \theta_i }{\kappa_D~ {\max_{i \in \Vc}}~\phi_i} \Big\}.
         \label{pf:cond_sigma}
     \end{align}
     The constraint \eqref{pf:cond_sigma} ensures that the Lyapunov function \eqref{pf:lyap} is positive definite.
     Next, let us take the time-derivative of the Lyapunov function \eqref{pf:lyap} along the trajectories of dynamics \eqref{sys_cl:x} with $\zeta(t) = 0$:
     \begin{align}
         \dot V\big(x(t)\big) = x(t)^\top \big( A^\top \bar{P} + \bar{P} A \big) x(t)
         =- x(t)^\top \bar{Q} x(t), \label{pf:V_dot}
     \end{align}
     where
     \begin{align}
         \bar{Q} = \ba{ccc}
         2\sigma M^{-1} \big( L + \Theta \big) ~~&~ 0 ~~&~ -\sigma M^{-1} \Phi \\
         0 & 2 \big( M^{-1}H - \sigma I  \big) & 0 \\
         -\sigma M^{-1} \Phi & 0 & 2\kappa_D^{-1} M^{-1} \Phi 
         \ea. \non
     \end{align}
     The constraint \eqref{pf:cond_sigma} also ensures that matrix $\bar{Q}$ is positive definite.
     This implies that $\dot V\big(x(t)\big)$ in \eqref{pf:V_dot} is negative definite and the matrix $A$ in \eqref{sys_cl:x} is Hurwitz.
     \QEDB
\end{pf}
Lemma~\ref{lem:A_hurwitz} enables us to have the following assumption.
\begin{assumption} \label{ass:zero_init}
    The  networked control system \eqref{sys_cl:x} is at its equilibrium $x_e = 0$ before being attacked.  \QET
\end{assumption}
\subsubsection{\textbf{Defender strategy:}}
At the chosen detection agent $d \in \Vc_{-\rho}$, the defender employs a detector as follow:
\begin{align}
    \dot{\hat{x}}_d(t) &= A \hat x_d(t) + K_d \eta_d(t),~~ \hat x_d(0) = 0, \label{obs:x_d} \\
    {\eta_d(t)} &=  y_d(t) - C_d \hat x_d(t), \label{obs:y_r}
\end{align}
where $\hat x_d(t) \in \Rbb^N$ is the estimated state of the networked system observed at agent $d$ and $\eta_d(t) \in \Rbb$ is the residual signal which will be used to detect cyber-attacks.
Since the result in \textit{Lemma~\ref{lem:A_hurwitz}} implies that $(A,C_d)$ is detectable,
matrix $K_d$ can be suitably designed such that the matrix $\big( A - K_d C_d \big)$ is Hurwitz.
Let us denote 
$\tilde x_d(t) \triangleq x(t) - \hat x_d(t)$ and $z_d(t) \triangleq \left[ x(t)^\top,~\tilde x_d(t)^\top \right]^\top$. From \eqref{sys_cl:x}-\eqref{sys_cl:yrho} and \eqref{obs:x_d}-\eqref{obs:y_r},
the augmented model can be rewritten as follows:
\begin{align}
    \dot z_d(t) &= A_d z_d(t) + \bar{E}_a \zeta(t), \label{sys_cl:zd} \\
    y_\rho(t) &= \bar{C}_\rho z_d(t), \label{sys_cl:yp} \\
    \eta_d(t) &= \bar C_d z_d(t), \label{sys_cl:yr}
\end{align}
where $y_\rho(t)$ and $\eta_d(t)$ are the outputs of the protected performance agent $\rho$ and the residual signal generated by the detector placed at agent $d \in \Vc_{-\rho}$, respectively; and
\begin{align}
    A_d &= \ba{cc}
    A & ~~ 0 \\
    0 & ~~ A - K_d C_d
    \ea, ~ \bar{E}_a = \ba{c}
    E_a \\ E_a
    \ea, \non \\
    \bar C_\rho &= \ba{cc}
    C_\rho &~~0^\top
    \ea, ~
    \bar C_d = \ba{cc}
    0^\top &~~ C_d
    \ea.
    \label{sys_cl:matr_ABCD}
\end{align}
We suppose that
the defender detects cyber-attacks if the energy of the residual signal over a given time horizon $[0,T]$ exceeds a given threshold $\delta$, i.e., $\norm{\eta_d}_{\Lc_{2}[0,T]}^2 > \delta^2$.
\subsubsection{\textbf{Adversary strategy:}}
The goal of the adversary is to maximally disrupt the output of the protected performance agent $\rho$ while remaining stealthy to the detector placed at agent $d$.
To this end, the adversary conducts the stealthy data injection attack, which is defined as follows.
Consider the continuous-time system \eqref{sys_cl:zd}, \eqref{sys_cl:yr},
the attack input signal $\zeta(t)$ 
is called the stealthy data injection attack if the residual signal satisfies $\norm{\eta_d}_{\Lc_{2}}^2 \leq \delta^2$ {where $\delta > 0$ is given and called an alarm threshold}. 
\subsection{Worst-case attack impact on the local performance}
Consider the continuous-time system \eqref{sys_cl:zd}-\eqref{sys_cl:yr} denoted as $\Sigma_{\rho d} \triangleq (A_d,\bar{E}_a,[\bar{C}_\rho^\top,\bar{C}_d^\top]^\top,0)$.
The malicious adversary attacks the input of the attack agent $a$ with the purpose of maliciously maximizing impact on the output of the given performance agent $\rho$ while remaining undetected by the defender.
This adversary purpose is translated into the following non-convex optimal control problem \citep[Sec. 4]{teixeira2021security}:
\begin{align}
    \gamma_\rho^\star(a,d) \triangleq \sup_{\zeta \in \Lc_{2e}, 
    z_d(0) = 0} &\norm{y_\rho}^2_{\Lc_{2}} \label{gamma} \\
    \text{s.t.} ~~~~~~& \norm{\eta_d}_{\Lc_{2}}^2 \leq \delta^2,
    \non
\end{align}
{
which has the dual problem as follows:
\begin{align}
    \inf_{\gamma_\rho \in \Rbb_+} \Big[ ~\sup_{\zeta \in \Lc_{2e}, 
    z_d(0) = 0} \big( \norm{y_\rho}^2_{\Lc_{2}} - \gamma_\rho  \delta^{-2} \norm{\eta_d}_{\Lc_{2}}^2  \big)  + \gamma_\rho\Big]. \label{gamma_dual}
\end{align}
The dual problem \eqref{gamma_dual} is feasible if $ \norm{y_\rho}^2_{\Lc_{2}} - \gamma_\rho  \delta^{-2} \norm{\eta_d}_{\Lc_{2}}^2 \allowbreak  \leq 0,~ \forall \zeta \in \Lc_{2e}$ and $z_d(0) = 0$, 
which results in the following optimization problem:
}
\begin{align}
    \gamma_\rho^\star(a,d) \triangleq \min_{\gamma_\rho \in \Rbb_+} ~~& \gamma_\rho \label{gamma_p} \\
    \text{s.t.} ~~~& \norm{y_\rho}^2_{\Lc_{2}} \leq {\gamma_\rho}{\delta^{-2}}  \norm{\eta_d}^2_{\Lc_{2}},~ \forall \zeta \in \Lc_{2e}, \non \\
    &~~~~~~~~~~~~~~~~~~~~~~~~~~~~~~~~~~~
    z_d(0) = 0.
    \non 
\end{align}

The strong duality can be proven by utilizing S-Procedure \citep[Ch .4]{petersen2000robust}. Recalling the key results in dissipative system theory for linear systems with quadratic supply rates \citep{trentelman1991dissipation}, the constraint of \eqref{gamma_p} can be translated into a linear matrix inequality \citep[Prop. 1]{teixeira2021security} as follows:
\begin{align}
	\gamma_\rho^\star(a,d) \triangleq  \underset{\gamma_\rho \in \Rbb_+, F = F^\top \geq  0}{\min} & ~~~~ \gamma_\rho
	\label{opt_lmis}
	 \\
	\text{s.t.}  ~~~~~~ 
	&R \big( \Sigma_{\rho d}, F, \gamma_\rho
	\big) \leq 0,  
	\non  
\end{align}
where
\begin{align}
   R \big( \Sigma_{\rho d}, F, \gamma_\rho
	\big) \triangleq & 
	\ba{cc}
	A_d^\top F - F A_d ~&~ F \bar{E}_a
	\\ 
	\bar{E}_a^\top F ~&~ 0
	\ea 
	\non \\
	&- 
	\ba{cc}
	{\gamma_\rho} {\delta^{-2}} \bar C_d \bar C_d^\top - \bar{C}_\rho \bar{C}_\rho^\top ~&~ 0 
	\\ 
	0 ~&~ 0
	\ea. \non
\end{align}
{The convex optimization problem \eqref{opt_lmis} can be solved numerically efficiently to obtain the worst-case attack impact on the local performance measured at the performance agent $\rho$.
With this worst-case attack impact, 
}
we are ready to state the following problem that will be addressed in the remainder of the paper.
\begin{problem}
\label{prob:det_selection}
    Given a protected performance agent $\rho$ and an arbitrary attack agent $a\in \Vc_{-\rho}$, select a detection agent $d \in \Vc_{-\rho}$ on which to place a detector that minimizes the worst-case attack impact on the performance agent $\rho$.
\end{problem}
{
\begin{remark}
    The two strategic players, which are the adversary and the defender, have symmetric information as described in Section~\ref{subsec:resource}. They  know the action space of their competitors instead of actual actions. Therefore, we assume that the two players perform their actions based on such available information at the same time, resulting in a non-cooperative game \citep{bacsar1998dynamic} which will be presented in the following section.
\end{remark}
}

\section{Optimal detector placement}
\label{sec:opt_dectector}
We first present a necessary and sufficient condition for the defender to ensure that the worst-case attack impact on the local performance is bounded.
This condition restricts the possible choices of the defender to a subset of available agents.
Then, we translate \textit{Problem~\ref{prob:det_selection}} into a zero-sum game between two strategic players, namely the malicious adversary and the defender.
Finally, within the framework of zero-sum games,
an algorithm is proposed to find Nash equilibria that yield the best strategies for the two players.

\subsection{Boundedness of the worst-case attack impact on the local performance}
Let us evaluate the attack impact of the adversary through the optimization problem \eqref{gamma}.
The feasibility of the optimization problem \eqref{gamma} is related to invariant zeros of $\Sigma_{\rho} = \big( A_d,\bar{E}_a,\bar{C}_\rho,0 \big)$ and $\Sigma_{d} = \big( A_d,\bar{E}_a,\bar{C}_d,0 \big)$, which are defined as follows.
\begin{definition} \label{def:invariant_zero}
	(Invariant zeros)
	Consider the strictly proper system $\bar \Sigma \triangleq (\bar A,\bar B,\bar C,0)$ with $\bar A,\bar B,$ and $\bar C$ are real matrices with appropriate dimensions. A tuple $(\bar \lambda,\bar{x},\bar g) \in \Cbb \times \Cbb^N \times \Cbb$ is a zero dynamics of $\Sigma$ if it satisfies
	\begin{align}
		\ba{cc}
		\lambda I - \bar A ~~~~ & -\bar B \\
		\bar C & 0
		\ea
		\ba{c}
		\bar{x} \\ \bar g
		\ea
		=
		\ba{c}
		0 \\ 0
		\ea,
		~~~ \bar{x} \neq 0.
		\label{definv:mtr_pen}
	\end{align}
	In this case, a finite $\bar \lambda$ is called a finite invariant zero of $\bar \Sigma$.
	Further, the strictly proper system $\bar \Sigma$ always has at least one invariant zero at infinity \citep{franklin2002feedback}.
	\QET
\end{definition}
More specifically, let us state the following lemma.
\begin{lemma} \label{lem:opt_feasible} \cite[Th. 2]{teixeira2015strategic}
    Consider the two following continuous time systems $\Sigma_{\rho} = \big( A_d,\bar{E}_a,\bar{C}_\rho,0 \big)$ and $\Sigma_d = \big( A_d,\bar{E}_a,\bar{C}_d,0 \big)$.
    The optimization problem \eqref{gamma} is feasible if and only if the unstable invariant zeros of $\Sigma_d$ are also invariant zeros of $\Sigma_{\rho}$.
    \QET
\end{lemma}
Inspired by the result in \textit{Lemma~\ref{lem:opt_feasible}} and the definition of invariant zeros in \textit{Definition~\ref{def:invariant_zero}}, we will investigate both finite and infinite invariant zeros of the two systems $\Sigma_d$ and $\Sigma_\rho$.
\subsubsection{\textbf{Finite invariant zeros:}}
Let us state the following lemma that considers the finite invariant zeros.
\begin{lemma} \label{lem:finite_zero_Sig_d}
    Consider the system {$\Sigma_m = (A,E_a,C_d,0)$} defined in~\eqref{sys_cl:x},~\eqref{sys_cl:yi} and, for $\lambda_d\in\mathbb{C}$, define \begin{align}
        \Qc(\lambda_d) &= L + \Theta + \lambda_d^2 M + \lambda_d H + \frac{\lambda_d \kappa_D}{\tau \lambda_d + 1} \Phi. \label{pf:Q_der} 
    \end{align}
    The system $\Sigma_m$ has a finite zero at $\lambda_d \in\mathbb{C}$ if, and only if, $\Qc(\lambda_d)$ is non-singular and $e_d^\top \Qc(\lambda_d)^{-1} e_a = 0$ \QET 
\end{lemma}
\begin{pf}
    The proof is postponed to Appendix~A.
    \QEDB
\end{pf}
The above result establishes the equivalence between the existence of an invariant zero of $\Sigma_m$ at $\lambda_d\in\mathbb{C}$ and the matrix $\Qc(\lambda_d)^{-1}$ having a zero at the entry $\left[\Qc(\lambda_d)^{-1}\right]_{da}$. Next, we leverage this result to show that the detector $\Sigma_d = \big( A_d,\bar{E}_a,\bar{C}_d,0 \big)$ has no unstable zero on the real line.
\begin{lemma} \label{lem:finite_zero}
     Consider system dynamics \eqref{sys_cl:zd},\eqref{sys_cl:yr} represented by $\Sigma_d = \big( A_d,\bar{E}_a,\bar{C}_d,0 \big)$ and assume that $\Gc$ is an undirected connected graph.
     Then, for any choice of attack agent $a\in \Vc_{-\rho}$ and detection agent $d \in \Vc_{-\rho}$,
     the corresponding system $\Sigma_d$ has no finite invariant zero on the positive real line.
     \QET
\end{lemma}
\begin{pf}
    The proof is postponed to Appendix~B.
    \QEDB
\end{pf}
Unfortunately, the result in \textit{Lemma~\ref{lem:finite_zero}} cannot be directly extended to consider complex invariant zeros on the right half-plane. 
The extension on how to deal with complex invariant zeros is left for future work. In the remainder of the paper, we assume that the system $\Sigma_d$ has no finite, complex unstable zeros.
\subsubsection{\textbf{Infinite invariant zeros:}}
We now investigate the infinite invariant zeros of the systems $\Sigma_\rho$ and $\Sigma_d$. 
In the investigation, we make use of known results connecting infinite invariant zeros and the relative degree (see \citet[Ch. 13]{khalil2002nonlinear}) of a linear system.
Let us denote $r_{(\rho,a)}$ and $r_{(d,a)}$ as the relative degrees of $\Sigma_\rho$ and $\Sigma_d$, respectively.
By following our existing result related to those infinite zeros \citep[Th. 7]{nguyen2022single},
the infinite zeros of $\Sigma_d$ are also the infinite zeros of $\Sigma_\rho$ if and only if the following condition holds
\begin{align}
     r_{(d,a)} \leq r_{(\rho, a)}.
     \label{cond:rel_deg}
\end{align}
The following theorem presents the necessary and sufficient condition which ensures that the optimization problem \eqref{gamma} admits a finite solution.
\begin{theorem} \label{th:feasibility}
    Consider a networked control system associated with an undirected connected graph $\Gc = \big( \Vc,\Ec, \Ac \big)$ and two continuous-time systems $\Sigma_\rho = \big(A_d,\bar{E}_a,\bar{C}_\rho,0\big)$ and $\Sigma_d =\big(A_d,\bar{E}_a,\bar{C}_d,0\big)$.
    Suppose $\Sigma_\rho$ and $\Sigma_d$ have relative degrees $r_{(\rho,a)}$ and $r_{(d,a)}$, respectively.
    The optimization problem \eqref{gamma} admits a finite solution if, and only if, the condition \eqref{cond:rel_deg} holds and the parameters $\theta_i,~\phi_i,~ \kappa_D,\text{and}~\tau \in \Rbb_+$ are such that, for every $\lambda_d\in\mathbb{C}$ on the right half plane, the matrix $\Qc(\lambda_d)^{-1}$ has no zero entries. 
    \QET
\end{theorem}
\begin{pf}
    Following from \textit{Lemma~\ref{lem:finite_zero_Sig_d}}, a suitable choice of parameters $\theta_i,~\phi_i,~ \kappa_D,\text{and}~\tau \in \Rbb_+$ ensures that the system $\Sigma_d$ has no finite unstable zero for any choice of $a$ and $d$ if, and only if, the matrix $\Qc(\lambda_d)^{-1}$ has no zero entries for every $\lambda_d\in\mathbb{C}$ on the right half plane.
    This result and the condition \eqref{cond:rel_deg} fulfill the necessary and sufficient condition in \textit{Lemma~\ref{lem:opt_feasible}} which guarantees that the optimization problem \eqref{gamma} admits a finite solution.
    \QEDB
\end{pf}
For every arbitrary attack agent $a \in \Vc_{-\rho}$,
let us define the detection set $\Dc \subseteq \Vc_{-\rho}$ containing agents which satisfy the necessary and sufficient condition in \textit{Theorem~\ref{th:feasibility}}. 
The possible action set of the defender will be restricted to the detection set $\Dc$.
\begin{assumption}
\label{assumption:dection}
    The detection set $\Dc$ is not empty, i.e., $\Dc = \{d_1,d_2,\ldots,d_{|\Dc|} \}$ where $|\Dc| \geq 1$.
    \QET
\end{assumption}
\textit{Assumption~\ref{assumption:dection}} enables the defender to optimally select an agent on which to place the observer \eqref{obs:x_d}-\eqref{obs:y_r} with the purpose of detecting the cyber-attack conducted by the adversary.
How the defender selects the optimal detector placement will be addressed by a game-theoretic approach, which has been widely used in \citet{pirani2021game,van2018distributed}, in the next subsection.
\begin{remark}
    \label{remark:find_dec_set} To compute a detection set $\Dc$ for a
    given undirected connected graph $\Gc$, we can utilize an undirected unweighted graph $\Gc'$ such that $\Gc$ and $\Gc'$ have the same topology. Through the graph $\Gc'$, we adopt the result in \citet[Lem. 8]{nguyen2022single} to  characterize candidate detection agents that fulfill the condition \eqref{cond:rel_deg} for every attack agent $a \in \Vc_{-\rho}$. Such found agents also satisfy the condition \eqref{cond:rel_deg} for every attack agent $a \in \Vc_{-\rho}$ in case we consider $\Gc$.
\end{remark}
\subsection{Game-theoretic approach to optimal detector placement}
According to \textit{Theorem~\ref{th:feasibility}},
since the optimization problem \eqref{gamma} is feasible for all the possible choices of the attack agent $a \in \Vc_{-\rho}$ and the detection agent $d \in \Dc$, we  employ the worst-case attack impact \eqref{gamma} as a game payoff that enables us to translate \textit{Problem 1} into a zero-sum game between the malicious adversary and the defender.
While the adversary wants to maximize the game payoff, the defender desires to minimize the same game payoff, i.e.,
\textit{Problem~\ref{prob:det_selection}} is represented as follows
\begin{align}
    \max_{a \in \Vc_{-\rho}}~~ \min_{d \in \Dc} ~~ \gamma_\rho^\star (a,d) < \infty.
    \label{min-max}
\end{align}
For every pair of an attack agent $a \in \Vc_{-\rho}$ and a detection agent $d \in \Dc$, we find the corresponding game payoff $\gamma^\star_\rho(a,d)$ by solving the convex optimization problem \eqref{opt_lmis}.
{
Then, the existence of a pure Nash equilibrium $(a^\star,d^\star)$ is equivalent to concluding that the following equality holds}
\begin{align}
    \min_{d_i \in \Dc} \big[ \alpha_i \big] = \max_{a_i \in \Vc_{-\rho}} \big[ \beta_i \big],
    \label{game_NE_checking}
\end{align}
where
$\alpha_i = \max_{a_j \in \Vc_{-\rho}} \gamma_\rho^\star(a_j,d_i)$;  $\beta_i = \min_{d_j \in \Dc} \gamma_\rho^\star (a_i,d_j)$.
The pure optimal detector placement at the detection agent $d_i^\star$ has the same index $i$ with $\alpha^\star_i$ where
\begin{align}
    \alpha^\star_i = \arg \min_{d_i \in \Dc}~ \big[ \alpha_i \big]. \label{game_find_d}
\end{align}
The failure of the condition \eqref{game_NE_checking} implies that no
pure Nash equilibrium exists \citep{zhu2015game}.
However, the game always admits a mixed-strategy Nash equilibrium which will be computed in the remainder of this section.
\begin{algorithm} [t]
\caption{Optimal detector placement \label{alg:NE_det}}
\begin{algorithmic}[1]
    \Statex{{\bf Input:} 
    possible detection set $\Dc$ and attack set $\Vc_{-\rho}$.
    }
    \Statex{{\bf Output:} 
    optimal detector placement
    }
    \State
    For every pair of $a \in \Vc_{-\rho}$ and $d \in \Dc$, solve \eqref{opt_lmis} to obtain the corresponding game payoff $\gamma_\rho^\star(a,d)$.
    \If{condition \eqref{game_NE_checking} is fulfilled} 
    \Statex{ \hskip0.4cm
    \multiline{
    \Return
    a pure detector placement at $ d^\star_i$ where its index $i$ is determined by \eqref{game_find_d}.
    }
    }
    \Else{} {solve \eqref{exp_payoff} to obtain  $P^\star$ and $Q^\star$}
    \Statex{ \hskip0.4cm 
    \multiline{
    \Return
    a mixed-strategy optimal detector placement represented by $Q^\star$.
    }
    }
    \EndIf
\end{algorithmic}
\end{algorithm}

Let us denote the probability of an agent $a \in \Vc_{-\rho}$ that is attacked by the adversary as $\pfr_a \in \Rbb_{[0;1]}$; the probability of an agent $d \in \Dc$ that is employed to implement the detector \eqref{obs:x_d}-\eqref{obs:y_r} by the defender as $\qfr_d \in \Rbb_{[0;1]}$
; vectors $P = [\pfr_i]$ and $Q = [ \qfr_i ]$.
According to \citet{zhu2015game},
the optimal mixed-strategy $(P^\star,Q^\star)$ of the adversary and the defender can be found as follows:
\begin{align}
     J_\rho^\star(P^\star,Q^\star) = \min_{P} ~ \max_{Q}~& \sum_{a \in \Vc_{-\rho}}~ \sum_{d \in \Dc} \pfr_a \gamma^\star_p(a,d) \qfr_d, \label{exp_payoff} \\
     \text{s.t.}~& \sum_{a \in \Vc_{-\rho}} \pfr_a =1,~~
    \sum_{d \in \Dc}\qfr_d =1, \non 
\end{align}
Inspired by \citet[Ch. 5]{boyd2004convex},
the min-max optimization problem \eqref{exp_payoff} can be efficiently solved by linear programming.
Let us summarize the procedure how to determine the optimal detector placement in \textit{Algorithm~\ref{alg:NE_det}}.
In the following section, we will demonstrate our proposed \textit{Algorithm~\ref{alg:NE_det}} in a case study of power systems.
\section{A case study}
\begin{figure} [!t]
    \centering
    \includegraphics[width=\linewidth]{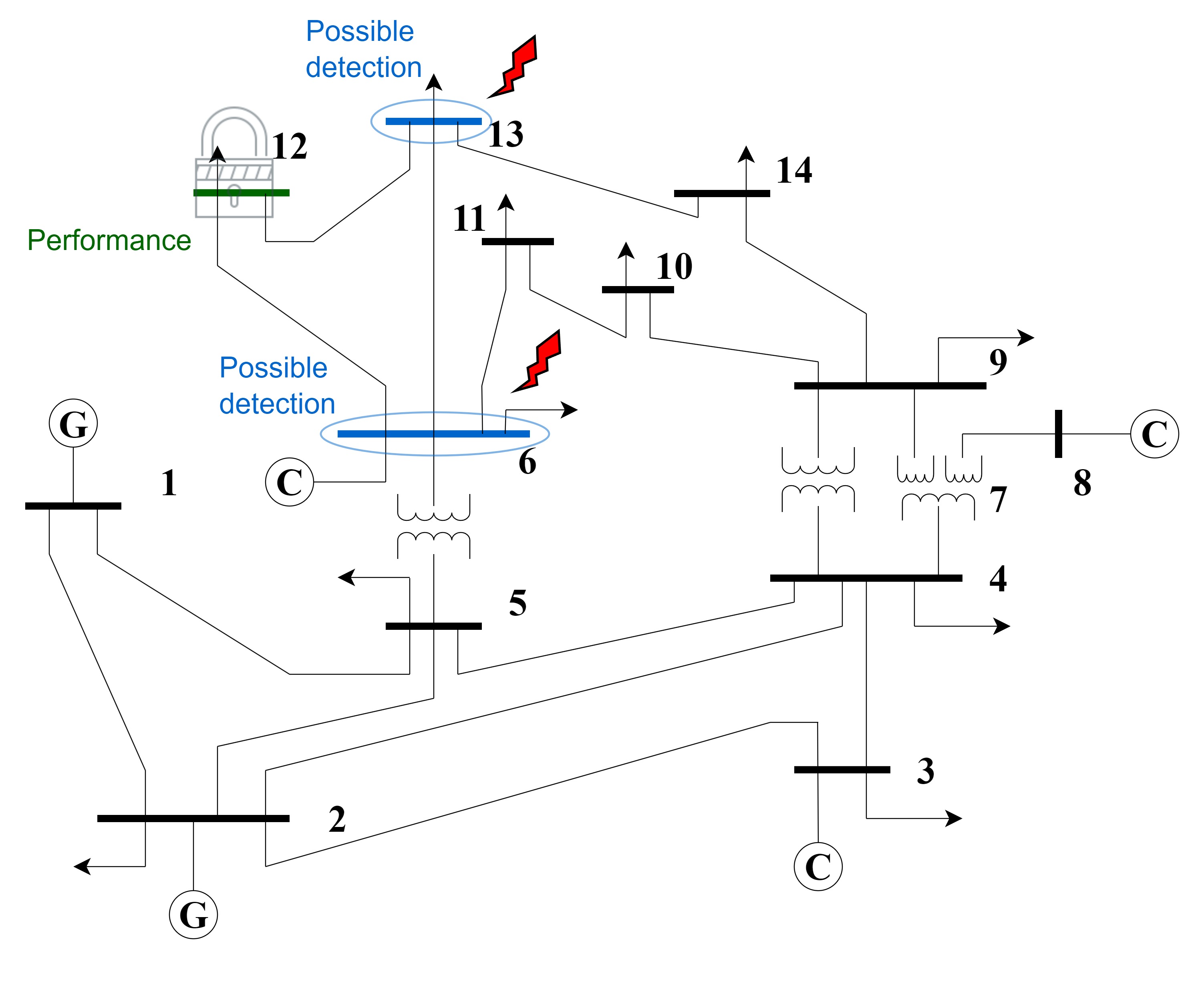}
    \caption{IEEE 14-bus system where bus $12$ (green) is the protected performance bus, buses $6$ and $13$ (blue) are possible detection buses. Buses $6~(56.2\%)$ and $13~(43.8\%)$ are possibly attacked.}
    \label{fig:ieee_14_bus}
\end{figure}

In this section, we demonstrate our obtained results via the IEEE 14-bus system (Fig.~\ref{fig:ieee_14_bus}).
The system includes 14 buses and 20 transmission lines.
The behavior of a bus $i \in \{1,2,\ldots,14\}$ can be described by the so-called swing equation \citep{tegling2018fundamental}:
\begin{align}
    m_i \ddot p_i + h_i \dot p_i - \tilde u_i(t) = -\sum_{j \in \Nc_i} Po_{ij}, \label{bus_dyn}
\end{align}
where $m_i$ and $h_i$ are the inertia and damping coefficients, respectively, $\tilde u_i(t)$ is the healthy/attacked mechanical input power and $Po_{ij}$ is the active power flow from bus $j$ to bus $i$. 
Considering that there are no power losses and $V_i = |V_i|e^{j p_i}~(j^2 = -1)$ and $p_i$ be the complex voltage and the phase angle of the bus $i$, respectively. 
The active power flow $Po_{ij}$ from bus $j$ to bus $i$ is given by 
\begin{align}
    Po_{ij} = -\ell_{ij} \sin(p_i - p_j), \label{power_ij}
\end{align}
where $-\ell_{ij} \in \Rbb_+$ is the susceptance of the power transmission line connecting bus $i$ with bus $j$.
Those parameters consisting of line susceptance $-\ell_{ij}$, inertia $m_i$, and damping $h_i$ can be found at \citet{ieee14bus}.
Since the phase angles usually are close, we can linearize \eqref{power_ij} and rewrite the dynamics \eqref{bus_dyn} of bus $i$ as follows
\begin{align}
    m_i \ddot p_i + h_i \dot p_i =  \sum_{j \in \Nc_i} \ell_{ij} \Big( p_i(t) - p_j(t) \Big) + \tilde u_i(t),
\end{align}
which is equivalent to the ones in \eqref{sys:pi}-\eqref{sys:yi} we investigated in the previous sections.
Suppose that the mechanic power input $\tilde u_i(t)$ coincides with the one in \eqref{sys:ui}-\eqref{sys:uai}.

Next, we present numerical results by using \textit{Algorithm~\ref{alg:NE_det}}.
Suppose that bus $12$ (coded green) is the protected performance bus.
The certain alarm threshold is selected as $\delta^2 = 2.6$.
{Recalling \textit{Remark~\ref{remark:find_dec_set}}}, we characterize the possible detection set $\Dc = \{6,~13\}$ containing buses that fulfill the condition \eqref{cond:rel_deg}.
The control parameters are selected as follows: $\theta_i = 1.5$, $\phi_i = 2.2$, $\kappa_d = 2$, and $\tau = 0.4$ $\forall i \in \Vc$.
Those control parameters fulfill the necessary and sufficient condition in \textit{Theorem~\ref{th:feasibility}} to ensure that the game payoff is bounded.
At the \textit{step 1} of \textit{Algorithm~\ref{alg:NE_det}},
for every pair of $a \in \Vc_{-\rho}$ and $d \in \Dc$, we solve \eqref{opt_lmis} by using CVX \citep{grant2014cvx} to obtain the following result: 
$\alpha = \big[4.7449,~4.3917\big]$ and $\beta = \big[2.4494,~2.5561,~2.6185,~2.5585,~2.4198,~2.3087,~2.5199,\allowbreak ~2.5257,~2.4695,~2.4673,~2.3705,~2.0717,~2.2119 \big]$.
At the \textit{step 2} of \textit{Algorithm~\ref{alg:NE_det}},
since $4.3917 = \min [\alpha_i] \neq \max [\beta_i] = 2.6185$, the condition \eqref{game_NE_checking} fails, implying that the zero-sum game does not admit a pure Nash equilibrium.
Then, we move to the \textit{step 3} to find a mixed-strategy $J_\rho^\star(P^\star,Q^\star) = 3.3757$ at $\pfr^\star_{6} = 0.562$, $\pfr^\star_{13} = 0.438$, $\pfr^\star_{i \in \Vc \setminus \{6,~12,~13\}} = 0$, $\qfr^\star_6 = 0.4878$, and $\qfr^\star_{13} = 0.5122$.

\begin{figure} [!h]
	\begin{center}
		\begin{subfigure}{.4\textwidth}
			\centering
			\includegraphics[width=\textwidth]{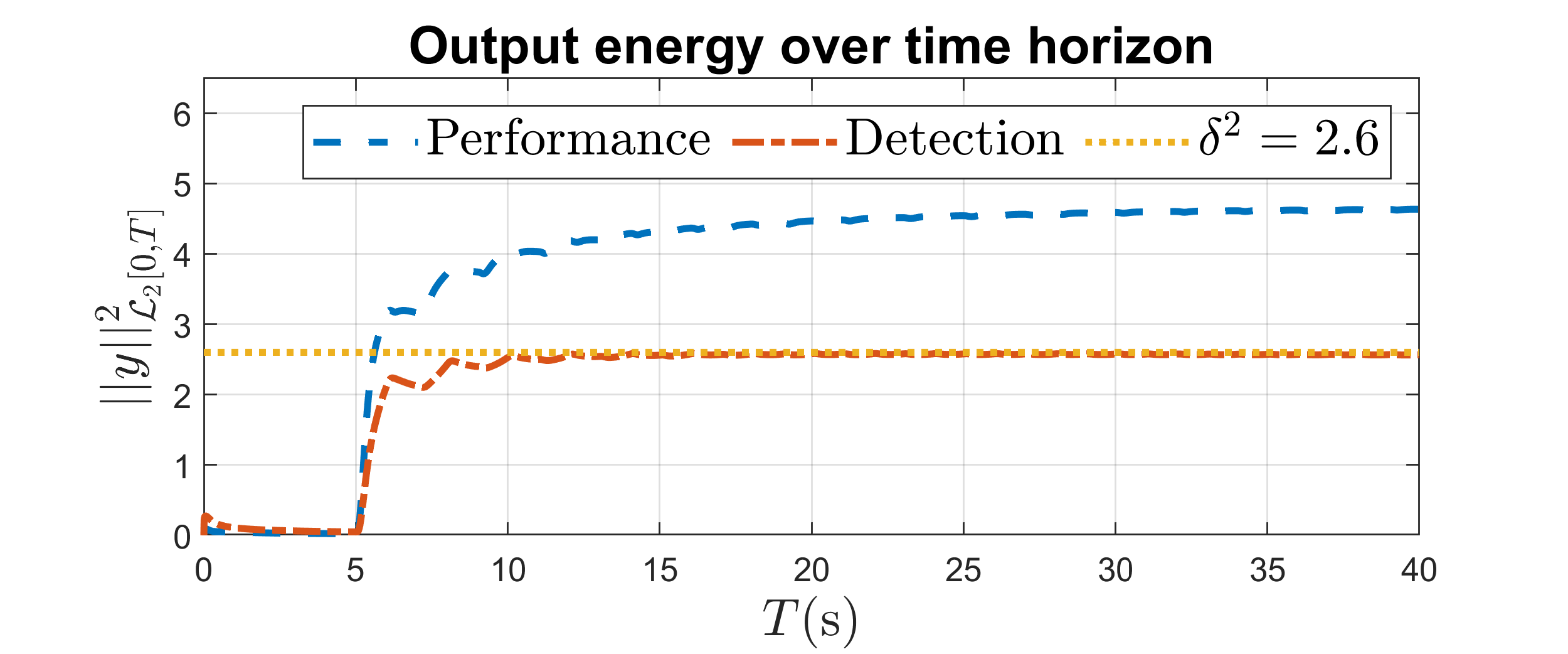}
			\caption{}
			\label{fig:energy}
		\end{subfigure}
		\begin{subfigure}{.4\textwidth}
			\centering
			\includegraphics[width=\textwidth]{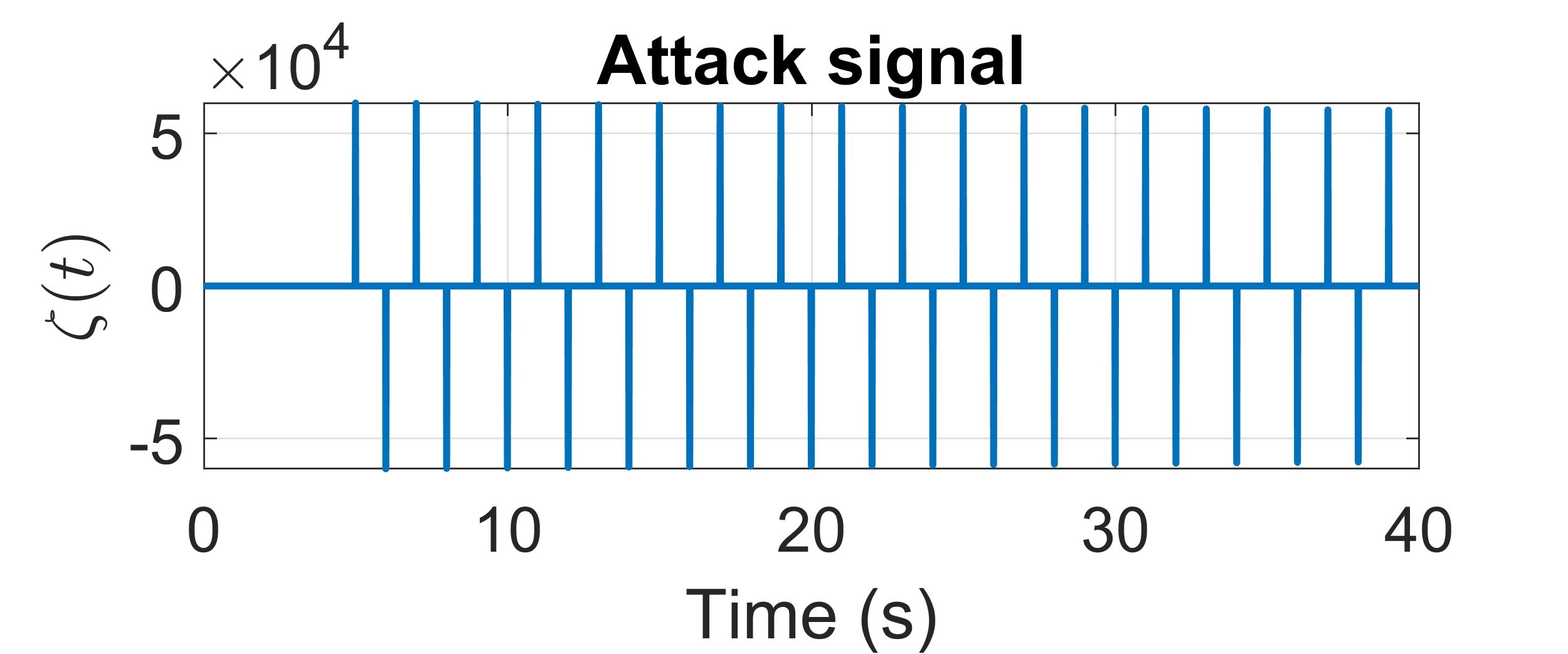}
			\caption{}
			\label{fig:attack}
		\end{subfigure}
	\end{center}	
	\caption{(a) Output energy of the performance bus $12$ and the detection bus $13$; (b) Attack signal $\zeta(t)$ conducted on the input of bus $6$.
	\label{fig:energy_attack}}
\end{figure}

Let us assume that the defender places a detector at the local controller of bus $13$ and the adversary conducts the stealthy data injection attack on the input of bus $6$.
By observing the output energy of the detection bus $13$ in Fig.~\ref{fig:energy} which is under the certain threshold $\delta^2$, the attack signal in Fig.~\ref{fig:attack} is stealthy to the detector placed at bus $13$.
However, the adversary only causes a bounded malicious attack impact on the output energy of the local performance bus $12$ (see Fig.~\ref{fig:energy}).
The adversary cannot increase the amplitude of the attack signal to gain its attack impact on the output energy of the performance bus $12$ since the energy output of the detection bus $13$ crosses the certain threshold $\delta^2 = 2.6$, which enables the defender to detect the cyber-attack.

\section{Conclusions}
In this paper, we addressed the problem of optimal detector placement in a networked control system under cyber-attacks.
First,
we presented the necessary and sufficient condition, which is related to the suitable choice of control parameters and the relative degree of dynamic systems, to ensure that the worst-case attack impact on the local performance is bounded.
This condition restricts possible detection agents to a subset of available agents.
Then,
the problem of optimal detector placement was formulated as a zero-sum game between the defender and the adversary where the game payoff was represented by the bounded worst-case attack impact on the local performance. 
Finally, an algorithm was devoted to finding the optimal detector placement.
The obtained results were illustrated by an actual case study of power systems, namely the IEEE 14-bus system.

\section*{Appendix~A: Proof of Lemma~\ref{lem:finite_zero_Sig_d}}
Let us denote a tuple $\big(\lambda_d,\bar{x}_d,g_d \big) \in \Cbb \times \Cbb^{3N} \times \Cbb$ as a zero dynamics of $\Sigma_m$ where $\lambda_d$ is a finite invariant zero of $\Sigma_m$ and 
$\bar{x}_d = \left[ \nu_1^\top,\nu_2^\top,\nu_3^\top \right]^\top$ where $\nu_1,\nu_2,\nu_3 \in \Cbb^N$.
From the condition \eqref{definv:mtr_pen} in \textit{Definition~\ref{def:invariant_zero}}, $\big(\lambda_d,\bar{x}_d,g_d \big)$ of $\Sigma_m$ satisfies
\begin{align}
        \ba{c:c:c:c}
        \lambda_d I & -I & 0 & 0 \\
        M^{-1}\big(L+\Theta\big) & \lambda I + M^{-1}H & -M^{-1}\Phi & e_a \\
        0 & \frac{\kappa_D}{\tau}I & \Big( \lambda + \frac{1}{\tau}  \Big) I & 0 \\
        e_d^\top & 0 & 0 & 0
        \ea 
        \ba{c}
        \nu_{1} \\
        \nu_{2} \\
        \nu_{3} \\ \bar{g}
        \ea = 0. \non 
\end{align}
Solving the above system of equations partially for $\nu_3$ and $\nu_2$, as functions of $\nu_1$, and then for $\nu_1$ as a function of $\bar{g}$ gives us the remaining equation
\begin{align}
         &e_d^\top M \Qc(\lambda_d)^{-1} \frac{\lambda_d \kappa_D}{\tau \lambda_d + 1} e_a\bar{g} = 0,  \label{pf:Q_der1} \\
         &\Qc(\lambda_d) = L +\Theta + \lambda_d^2 M + \lambda_d H + \frac{\lambda_d \kappa_D}{\tau \lambda_d + 1} \Phi. \non
\end{align}
From \eqref{pf:Q_der1}, given the positivity of the parameters $\theta_i,~\phi_i,~ \\ \kappa_D,  \text{and}~\tau \in \Rbb_+$, 
it follows that
$\big(\lambda_d,\bar{x}_d,g_d \big) \in \Cbb \times \Cbb^{3N} \times \Cbb$ is a zero dynamics of $\Sigma_m$ with $\bar{g}\neq 0$ if, and only if, $e_d^\top M \Qc(\lambda_d)^{-1}e_a = \left[\Qc(\lambda_d)^{-1}\right]_{da}=0$ where matrix $M$ is a diagonal positive definite matrix.


\section*{Appendix~B: Proof of Lemma~\ref{lem:finite_zero}}
Let us consider the continuous-time systems $\Sigma_{mo} = \big( A-K_dC_d, E_a, C_d, 0 \big)$, $\Sigma_m = \big( A, E_a, C_d, 0 \big)$, and $\Sigma_o = \big( A-K_dC_d,K_d,-C_d,1 \big) $. 
From the condition \eqref{definv:mtr_pen} and the structure of matrices \eqref{sys_cl:matr_ABCD}, the set of invariant zeros of the system $\Sigma_d$ is the union of the set of eigenvalues of matrix $A$ and the set of invariant zeros of the system $\Sigma_{mo}$.
Thanks to \textit{Lemma~\ref{lem:A_hurwitz}}, all the eigenvalues of matrix $A$ is stable.
It remains to investigate invariant zeros of the system $\Sigma_{mo}$.
On the other hand, we have the set of invariant zeros of the $\Sigma_{mo}$ is contained by the union of the set of invariant zeros of $\Sigma_o$ and the set of invariant zeros of $\Sigma_m$.
By following \textit{Definition~\ref{def:invariant_zero}}, the condition \eqref{definv:mtr_pen} gives us that the invariant zeros of the system $\Sigma_o$ coincides with eigenvalues of matrix $A$ in \eqref{sys_cl:x}, which are stable, no matter how the matrix $K_d$ in the observer \eqref{obs:x_d} is designed.
In the end, we only need to investigate invariant zeros of $\Sigma_m$.
The proof follows from a contradiction argument. Let us denote a tuple $\big(\lambda_d,\bar{x}_d,g_d \big) \in \Cbb \times \Cbb^{3N} \times \Cbb$ as a zero dynamics of $\Sigma_m$ where $\lambda_d$ is assumed to be real and positive.

For every real positive value $\lambda_d$, the matrix $\Qc(\lambda_d)$ in \eqref{pf:Q_der} is positive definite, yielding that $\Qc(\lambda_d)$ is non-singular and $-\Qc(\lambda_d)$ is Hurwitz.
Further, since matrix $\Qc(\lambda_d)$ represents a strongly connected graph $\Gc$ with added self-loops,
it is irreducible \citep[Ch. 6]{horn2012matrix}.
Obviously, $-\Qc(\lambda_d)$ is also a Metzler matrix.
According to \citet[Th. 10.3]{bullo2019lectures}, $\Qc(\lambda_d)^{-1}$ is a positive matrix whose all entries are real positive, that is, $\left[\Qc(\lambda_d)^{-1}\right]_{da}>0$ for all vertices $d$ and $a$. Following the result of \textit{Lemma~\ref{lem:finite_zero_Sig_d}}, we conclude that a real positive value $\lambda_d$ cannot be a zero of the system $\Sigma_m$, thus concluding the proof.

\bibliography{ref.bib}

\end{document}